\documentclass[12pt,aps,prd,floats,floatfix,showpacs,twoside,preprintnumbers,superscriptaddress,nofootinbib]{revtex4}
\usepackage[dvips]{graphicx,pstricks}
\usepackage{amsmath}
\usepackage{amssymb}
\usepackage{hyperref}
\usepackage{epsfig}
\usepackage{color}

\newcommand{\be}{\begin{equation}}
\newcommand{\ee}{\end{equation}}
\newcommand{\ba}{\begin{eqnarray}}
\newcommand{\ea}{\end{eqnarray}}

\def\roughly#1{\mathrel{\raise.3ex\hbox{$#1$\kern-.75em%
\lower1ex\hbox{$\sim$}}}}

\def\gsim{\roughly>}

\def\slashchar#1{\setbox0=\hbox{$#1$}  
   \dimen0=\wd0     
   \setbox1=\hbox{/} \dimen1=\wd1  
   \ifdim\dimen0>\dimen1   
      \rlap{\hbox to \dimen0{\hfil/\hfil}} 
      #1     
   \else     
      \rlap{\hbox to \dimen1{\hfil$#1$\hfil}} 
      /      
   \fi}

\makeatletter
\def\overbracket#1{\mathop{\vbox{\ialign{##\crcr\noalign{\kern3\p@}
\downbracketfill\crcr\noalign{\kern3\p@\nointerlineskip}
$\hfil\displaystyle{#1}\hfil$\crcr}}}\limits}
\def\underbracket#1{\mathop{\vtop{\ialign{##\crcr
$\hfil\displaystyle{#1}\hfil$\crcr\noalign{\kern3\p@\nointerlineskip}
\upbracketfill\crcr\noalign{\kern3\p@}}}}\limits}
\def\upbracketfill{$\m@th\makesm@sh{\llap{\vrule\@height3\p@\@width.7\p@}}%
\leaders\vrule\@height.7\p@\hfill
\makesm@sh{\rlap{\vrule\@height3\p@\@width.7\p@}}$}
\def\downbracketfill{$\m@th
\makesm@sh{\llap{\vrule\@height.7\p@\@depth2.3\p@\@width.7\p@}}%
\leaders\vrule\@height.7\p@\hfill
\makesm@sh{\rlap{\vrule\@height.7\p@\@depth2.3\p@\@width.7\p@}}$}
\makeatother

\begin{document}

\date{\today}

\preprint{ZTF-EP-15-05}

\title{An effective model for the QCD phase transitions at finite baryon density}

\author{S.~Beni\' c}
\affiliation{Physics Department, Faculty of Science, University of Zagreb, Zagreb 10000, Croatia}
\affiliation{Department of Physics, The University of Tokyo, 7-3-1 Hongo, Bunkyo-ku, Tokyo 113-0033, Japan}

\author{I.~Mishustin}
\affiliation{Frankfurt Institute for Advanced Studies, D-60438 Frankfurt am Main, Germany}
\affiliation{Kurchatov Institute, Russian Research Center, Akademika Kurchatova Sqr., Moscow, 123182, Russia}

\author{C.~Sasaki}
\affiliation{Frankfurt Institute for Advanced Studies, D-60438 Frankfurt am Main, Germany}
\affiliation{Institute of Theoretical Physics, University of Wroc\l aw, PL-50204 Wroc\l aw, Poland}

\begin{abstract}

We introduce an effective quark-meson-nucleon model for the QCD phase transitions at finite baryon density.
The nucleon and the quark degrees of freedom are described within a unified framework of a chiral linear sigma model.
The deconfinement transition is modeled through a simple modification of the distribution 
functions of nucleons and quarks, where an additional auxiliary field, the bag field, is introduced.
The bag field plays a key role in converting between the nucleon and the quark degrees of freedom.
The model predicts that the chiral and the deconfinement phase transitions are always separated.
Depending on the model parameters, the chiral transition occurs in the 
baryon density range of $(1.5-15.5)n_0$, while the deconfinement 
transition occurs above $5 n_0$, where $n_0$ is the saturation density.

\end{abstract}

\pacs{12.39.Fe,12.38.Mh,25.75.Nq}

\maketitle

\section{Introduction}

The basic problem of QCD thermodynamics is to understand 
the conversion from hadrons to quarks and gluons and how this is related to the underlying chiral and deconfinement transitions.
This problem has major relevance to the physics of heavy ion collisions and compact stars \cite{Fukushima:2013rx,Brambilla:2014jmp,Fukushima:2014pha,Buballa:2014jta}.
At present there are several proposals to effectively convert between
the hadronic and the quark-gluon phase, with partial success.
At finite $T$, quark degrees of freedom can be suppressed by the Polyakov loop \cite{Fukushima:2003fw,Ratti:2005jh,Roessner:2006xn}.
Mechanisms for excluding composite degrees of freedom are the spectral function method \cite{Yamazaki:2012ux,Wergieluk:2012gd,Jakovac:2013iua,Yamazaki:2013yua,Blaschke:2013zaa,Blaschke:2015nma} and the excluded volume method, for recent works see Refs.~\cite{Steinheimer:2011ea,Dexheimer:2014pea,Benic:2014jia,Satarov:2014voa}.

In this work we construct an effective quark-meson-nucleon model with two flavors for the QCD phase transitions at finite density.
We place several restrictions to our approach: first we must take into account both the nuclear and the quark degrees of freedom. 
Second, the model should respect global symmetries of QCD, i.~e. the chiral symmetry and the scale invariance.
Third, quark degrees of freedom must be excluded at the 
nuclear matter density, and nuclear degrees of freedom must be excluded at some high density.
We require that the model reproduces the nuclear matter ground state.
Finally, a unified description must encompass couplings of 
nucleons and quarks to the same bosonic mean-fields generated from a unique vacuum potential.

For nucleons, both chiral and scale invariance 
can be accommodated in the 
parity doublet model \cite{Detar:1988kn,Hatsuda:1988mv,Jido:1998av,Zschiesche:2006zj,Dexheimer:2007tn,Sasaki:2010bp,Gallas:2011qp} \cite{Sasaki:2011ff,Paeng:2011hy,Paeng:2013xya}.
The quark sector is described by a linear sigma model \cite{Sasaki:2011sd,Boeckel:2011yj} coupled 
to the dilaton \cite{Schechter:1980ak}.
We find that the small $\sigma$ mass as required for reasonable 
nuclear matter properties \cite{Zschiesche:2006zj} results in a shallow potential in the $\sigma$ direction.
As an immediate consequence, pure quark matter appears at baryon chemical potentials $\mu_B$ below the value of the vacuum nuclear mass.
In order to solve this problem we generalize the idea of statistical confinement 
from effective model studies at finite $T$ to finite densities.
While at finite $T$ the Polyakov loop is used to statistically suppress 
thermal quark fluctuations \cite{Fukushima:2003fw}, its extension to finite baryon 
chemical potential is problematic. In this work we use instead the concept of 
infrared confinement \cite{Ebert:1996vx,Berges:1998ha} in order to modify the Fermi-Dirac distributions of quarks.
We consider a simple model where the Fermi distribution of quarks is 
restricted to momenta above $b$, where $b$ is a new auxiliary field
in our model, which we name the \emph{bag} field.
The finite value of $b$ in the vacuum and at low temperatures and densities is guaranteed by a 
new phenomenological vacuum potential.
We fit the parameters of this potential to the QCD vacuum energy and by matching 
the pseudo-critical temperature for the chiral and deconfinement transition 
known from the lattice simulations at $N_f=2$ \cite{Ejiri:2000bw,Maezawa:2007fd}.
The model is compared to a Polyakov-quark-meson model by pointing out their similarities and differences.

Next, we introduce a generalization of the distribution functions for the 
nucleons in such a way that their Fermi surface is restricted only to low momenta \cite{Berges:1998ha}, i.~e. below 
some value $\alpha b$, where $\alpha$ is a new parameter.
This leads to a construction of a combined quark-meson-nucleon model.
We find that the minimization of the thermodynamic potential in the $b$ field 
acts to convert the nucleons to quarks as the baryon chemical potential is increased.  
One of the main consequences of this model is that the chiral and the deconfinement transitions at $T=0$ are {\it separated}.

This paper is organized as follows: in Sec.~\ref{sec:nuc} we introduce the nucleonic model and briefly consider the nuclear matter ground state.
Sec.~\ref{sec:qu} is devoted to the quark degrees of freedom.
Here we introduce the $b$-field potential and discuss statistical confinement of quarks.
The Sec.~\ref{sec:qmn} describes a combined quark-meson-nucleon model.
The main results of this paper are given in Sec.~\ref{sec:res}, while in the following Sec.~\ref{sec:conc} we make our conclusions.
In the Appendix we solve a simplified version of the quark-meson-nucleon model. 

\section{Nucleonic sector: parity doublet model with dilaton}
\label{sec:nuc}

We consider the $N_f=2$ 
nuclear parity doublet model within the mirror assignment of chiral symmetry
\cite{Detar:1988kn,Hatsuda:1988mv,Jido:1998av,Zschiesche:2006zj,Dexheimer:2007tn,Sasaki:2010bp,Gallas:2011qp} \cite{Sasaki:2011ff,Paeng:2011hy,Paeng:2013xya}.
We prefer to use a linear realization of the chiral symmetry which allows the description of chiral symmetry restoration.
The restoration of chiral symmetry in QCD dictates that hadrons of opposite parity become degenerate, but not necessarily massless. 
A finite chirally invariant mass is then modeled by a parity doublet model with mirror assignment.
Note that the possibility of a chirally invariant contribution to the nucleon mass was recently hinted by lattice calculations \cite{Glozman:2012fj}.
First lattice QCD simulations of nucleon parity partners at finite $T$ \cite{Aarts:2015mma} find that their masses become degenerate by crossing the QCD phase transition.

The nucleonic part of the model Lagrangian \cite{Zschiesche:2006zj,Sasaki:2010bp} coupled 
to the dilaton \cite{Sasaki:2011ff,Paeng:2011hy,Paeng:2013xya} is
\be
\begin{split}
\mathcal{L}_N &= i\bar{\Psi}_1\slashchar{\partial}\Psi_1+
 i\bar{\Psi}_2\slashchar{\partial}\Psi_2+ g_\chi \chi(\bar{\Psi}_1\gamma_5\Psi_2-\bar{\Psi}_2\gamma_5\Psi_1)\\
 &+g_1\bar{\Psi}_1(\sigma+i\gamma_5\boldsymbol{\tau}\cdot \boldsymbol{\pi})\Psi_1+g_2\bar{\Psi}_2(\sigma-i\gamma_5\boldsymbol{\tau}\cdot \boldsymbol{\pi})\Psi_2\\
&-g_\omega\bar{\Psi}_1\slashchar{\omega}\Psi_1
-g_\omega\bar{\Psi}_2\slashchar{\omega}\Psi_2~.
\end{split}
\ee
where $\Psi_{1,2}$ are the nuclear chiral partners.
The fermions $\Psi_{1,2}$ are coupled to the chiral fields $(\sigma,\boldsymbol{\pi})$, to the $\omega_\mu$ field 
and to the dilaton $\chi$.
The mass eigenstates are given as
\be
\begin{pmatrix}
  N_+  \\
  N_- 
 \end{pmatrix}
 = \frac{1}{\sqrt{2\cosh\delta}}
 \begin{pmatrix}
   e^{\delta/2} & \gamma_5 e^{-\delta/2} \\
   \gamma_5 e^{-\delta/2} & -e^{\delta/2}
  \end{pmatrix}
\begin{pmatrix}
  \Psi_1  \\
  \Psi_2 
 \end{pmatrix}
~,
\ee
where 
$$\sinh \delta = -\frac{g_1+g_2}{2g_\chi}\frac{\sigma}{\chi}~,$$
with masses
\be
m_{N_\pm} = \frac{1}{2}\left[\sqrt{(g_1+g_2)^2\sigma^2 + 4g_\chi^2\chi^2}\mp (g_1-g_2)\sigma\right]~.
\ee
The state $N_+$ is the nucleon $N(938)$ while $N_-$ is its parity partner conventionally identified with $N(1500)$.
The meson contribution is as follows
\be
\begin{split}
\mathcal{L}_M &= \frac{1}{2}(\partial_\mu \sigma)^2+\frac{1}{2}(\partial_\mu\boldsymbol{\pi})^2+\frac{1}{2}(\partial_\mu\chi)^2-\frac{1}{4}(\omega_{\mu\nu})^2\\
&-V_\sigma-V_\omega-V_\chi~,
\end{split}
\ee
where
\be
V_\sigma = \frac{\lambda}{4}\left(\sigma^2 + \boldsymbol{\pi}^2-\frac{\lambda_\chi}{\lambda}\chi^2\right)^2-\epsilon\sigma\chi^2~,
\label{eq:sigpot}
\ee
\be
V_\omega=-\frac{\lambda_\omega}{2}\chi^2 \omega_\mu^2~,
\label{eq:vom}
\ee
and
\be
V_\chi=\frac{B}{4}\left(\frac{\chi}{\chi_0}\right)^4
\left[\log\left(\frac{\chi}{\chi_0}\right)^4 - 1\right]~.
\ee
The total Lagrangian $\mathcal{L}_N + \mathcal{L}_M$ is 
chiral and scale invariant.
All the masses in the model are generated by the condensation of the dilaton field in the vacuum $\chi_0$.
We can fix $\lambda_\omega$ by $m_\omega^2 = \lambda_\omega \chi_0^2$, where $m_\omega = 783$ MeV.
The parameters $\lambda$, $\lambda_\chi$ and $\epsilon$ are related to the sigma and pion masses and the pion decay constant $f_\pi$ as
\be
\lambda = \frac{m_\sigma^2 - m_\pi^2}{2f_\pi^2}~, \quad
\lambda_\chi = \frac{m_\sigma^2-3m_\pi^2}{2\chi_0^2}~,
\quad
\epsilon = \frac{m_\pi^2 f_\pi}{\chi_0^2}~,
\ee
with $m_\pi = 138$ MeV and $f_\pi = 93$ MeV.
We take $m_+ = 938$ MeV, $m_- = 1500$ MeV \cite{Zschiesche:2006zj}.
The nuclear matter ground state can be obtained by fixing the parameters $m_\sigma$, $g_1$, $g_2$, $g_\chi$, $g_\omega$.
The dilaton potential $V_\chi$ is fixed by identifying the lowest glueball mass with the dilaton mass $m_\chi = 1700$ MeV \cite{Sexton:1995kd,Chen:2005mg}, and by fixing the value of the gluon condensate.
The conventional value of the gluon condensate
$\langle \frac{\alpha_s}{\pi} G_{\mu\nu} G^{\mu\nu}\rangle \simeq (331 \, {\rm MeV})^4$ \cite{Shifman:1978bx} is accompanied by large uncertainties, 
$\langle \frac{\alpha_s}{\pi} G_{\mu\nu} G^{\mu\nu}\rangle \simeq (300-600 \, {\rm MeV})^4$. For recent accounts see \cite{Narison:2011xe,Janowski:2014ppa,Dominguez:2014pga} and references therein.
By the relation for the trace anomaly this 
can be translated into the following range for the
QCD vacuum energy $\epsilon_{\rm vac} \simeq (193-386 \, {\rm MeV})^4$.
Assuming that the QCD vacuum energy is dominated by the dilaton potential, the gluonic bag constant is estimated to $B \simeq (273-546 \, {\rm MeV})^4$.
From the relation for the dilaton mass
\be
m_\chi^2 = \frac{\partial^2 V_\chi}{\partial \chi^2} = \frac{4B}{\chi_0^2}~,
\ee
we obtain $\chi_0 \simeq 87.79-351.17$ MeV.

Since the dilaton is 
heavy $m_\chi = 1700$ MeV, it practically does not influence the nuclear ground state and we can adopt the model parameters from \cite{Zschiesche:2006zj}.
This fixes $m_\sigma = 370.63$ MeV, $g_1 = 13.00$, $g_2 = 6.97$,
$g_\chi = 4.39$ and $g_\omega = 6.79$.
The corresponding thermodynamic potential in the mean-field approximation is
\be
\Omega = V_\sigma+V_\omega+V_\chi + \sum_{X=N_\pm}\Omega_X~,
\ee
\be
\Omega_X = \gamma_N \int\frac{d^3 p}{(2\pi)^3}
\left[T\log\left(1-f_X\right)+
T\log\left(1-\bar{f}_X\right)\right]~, 
\label{eq:omthn}
\ee
where the functions $f_X$ are the Fermi-Dirac distributions
$$f_X = \frac{1}{1+e^{\beta(E_X-\mu_N)}}~, \qquad
\bar{f}_X = \frac{1}{1+e^{\beta(E_X+\mu_N)}}~,$$
and $E_X = \sqrt{\mathbf{p}^2 + m_X^2}$, $\mu_N = \mu_B - g_\omega \omega$, with $\gamma_N = 2\times 2$ being the spin-isospin degeneracy factor.
We minimize the potential with respect to $\sigma$, $\chi$ and $\omega$
\be
\frac{\partial\Omega}{\partial \sigma} = -\lambda_\chi\chi^2\sigma+\lambda\sigma^3-\epsilon\chi^2+
\sum_{X=N_\pm}\frac{\partial m_X}{\partial\sigma}s_X=0~,
\label{eq:gaps}
\ee
\be
\frac{\partial\Omega}{\partial \omega} = -\lambda_\omega\chi^2\omega + g_\omega\sum_{X=N_\pm}\rho_X=0~,
\label{eq:gapom}
\ee
\be
\begin{split}
\frac{\partial\Omega}{\partial \chi} &= -\lambda_\chi\sigma^2\chi+\frac{\lambda_\chi^2}{\lambda}\chi^3-2\epsilon\sigma\chi-\lambda_\omega\chi\omega^2+B\frac{\chi^3}{\chi_0^4}\log\left(\frac{\chi}{\chi_0}\right)^4\\
&+\sum_{X=N_\pm} \frac{\partial m_X}{\partial\chi}s_X=0~,
\end{split}
\label{eq:gapch}
\ee
where the scalar and the baryon number densities are, respectively
\be
s_X = \gamma_N\int\frac{d^3 p}{(2\pi)^3}\frac{m_X}{E_X}(f_X + \bar{f}_X )~,
\label{eq:scaln}
\ee
and
\be
\rho_X = \gamma_N\int\frac{d^3 p}{(2\pi)^3}(f_X - \bar{f}_X )~.
\label{eq:densn}
\ee

As mentioned above, with the present parametrization, the contribution of the dilaton field to the ground state properties is numerically negligible.
One can check whether $\chi$ can influence the nuclear matter equation of state at all.
In principle, this is possible but with a lower dilaton mass, see \cite{Papazoglou:1996hf}.
With a mass of $m_\chi = 1700$ MeV used in this work the impact of a dilaton field, e.~g. on chiral restoration, is expected only at much higher densities.
We briefly discuss this possibility in Sec.~\ref{sec:res}.

\section{Quark sector: linear sigma model and statistical confinement}
\label{sec:qu}

\begin{figure}[t]
\begin{center}
\includegraphics[clip,scale=0.4]{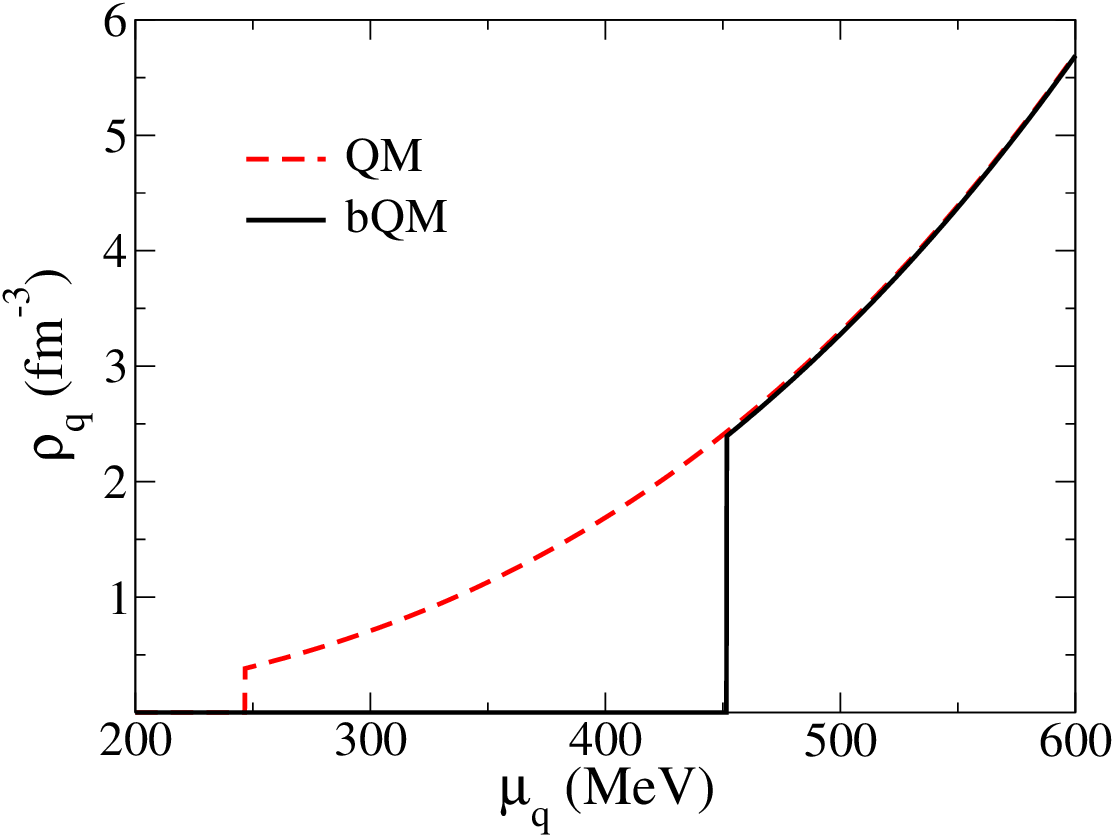}
\caption{Quark number density in the QM and in the bQM model.
}
\label{fig:densq}
\end{center}
\end{figure}
We introduce the quark-meson (QM) coupling according to the linear sigma model
\be
\mathcal{L}_q = i\bar{q}\slashchar{\partial}q 
+g_q \bar{q}(\sigma + i\gamma_5\boldsymbol{\tau}\cdot \boldsymbol{\pi})q-V_\sigma~,
\label{eq:lsm}
\ee
with the {\it same} parameters in the 
$\sigma$-potential as in Eq.~(\ref{eq:sigpot}) where we considered nuclear matter.
For the quark-meson coupling we use $g_q = 300 \, {\rm MeV}/f_\pi$.
The dilaton is too heavy and hence not essential for the following discussion, so we set $\chi = \chi_0$ by hand.
The thermodynamics of this model was studied in Refs.~\cite{Gocksch:1991up,Scavenius:2000qd,Bowman:2008kc}.

\subsection{Shallow potential and early onset of quarks}

Using the same model parameters  (i.~e. low $m_\sigma$) that are constrained in Sec.~\ref{sec:nuc} as to obtain reasonable nuclear matter properties
gives a shallow potential in the $\sigma$-direction.
A shallow potential leads to a chiral phase transition in cold quark matter at rather low densities, even below $n_0$ \cite{Scavenius:2000qd}.
The corresponding equation of state has a zero-pressure point where quark matter has a finite density.
This is illustrated by
the red line in Fig.~\ref{fig:densq} where we plot the density of quarks $\rho_q$
as a function of the quark chemical potential $\mu_q = \mu_B/3$.
In our present model we obtain $3\mu_q \simeq 750$ MeV which is lower than
the nucleon mass.

This is a striking problem: quarks appear too early due to a shallow potential.
A common way to solve this problem is to adjust the chiral potential in the quark sector independently of the nucleonic sector.
Such a treatment is not possible in a unified description of nucleon and quark matter.
Also, note that a larger $m_\sigma$ is not favored in view of nuclear matter ground state.
For example, with typical values in quark models $m_\sigma \gsim 600$ MeV, compressibility of nuclear matter at saturation increases by an order of magnitude from its experimentally suggested range \cite{Zschiesche:2006zj}.
In Ref.~\cite{Steinheimer:2011ea} the problem of a flat potential was circumvented by assigning a bare mass term of $200$ MeV to the quarks.
We conclude that the essential missing physics is confinement of quarks in the infrared region where their interaction becomes strong.
To avoid this inconsistency below we introduce a simple model of {\it statistical confinement}.

\subsection{Statistical confinement of quarks}

The concept of statistical confinement is very successful at finite temperatures where the Polyakov loop is used to modify the quark distribution functions \cite{Fukushima:2003fw}.
At finite density and small temperature the center symmetry is badly broken so we cannot use the Polyakov loop.
As an alternative, we propose a modification of the quark distribution functions via the following ansatz
\be
n_q = \theta(\mathbf{p}^2 - b^2) f_q~, \qquad \bar{n}_q = \theta(\mathbf{p}^2 - b^2) \bar{f}_q~,
\label{eq:ir}
\ee
where $b$ is a parameter, and $f_q$ and $\bar{f}_q$ are 
$$f_q = \frac{1}{1+e^{\beta(E_q-\mu_q)}}~, \qquad
\bar{f}_q = \frac{1}{1+e^{\beta(E_q+\mu_q)}}~,$$
the Fermi-Dirac distribution functions for quarks and antiquarks, respectively.
Obviously, in Eq.~(\ref{eq:ir}) quarks 
with momenta $\mathbf{p}^2<b^2$ are suppressed.

This is one possible way to restrict thermal quark fluctuations at low momenta.
It is similar to the concept of infrared confinement used in the Dyson-Schwinger 
vacuum studies \cite{Roberts:2010rn,Roberts:2011wy} and in the NJL model \cite{Ebert:1996vx,Bentz:2001vc,Blaschke:2001yj,Dubinin:2013yga}.
The infrared cutoff is in-line with the idea of in-hadron condensates
\cite{Brodsky:2008be},
and is also implemented in 
the holographic hard wall \cite{Polchinski:2001tt} and soft wall \cite{Karch:2006pv} models.
Intuitively, $1/b$ can be understood as a typical size of a hadron, 
so that due to the uncertainty principle quarks cannot have momenta lower than $b$.

With a sharp cutoff in the distribution function 
it is not possible to saturate the Stefan-Boltzmann limit at high temperature and/or density.
Essentially, $b$ must be a medium dependent quantity.
A thermodynamically consistent way to achieve this is to promote $b$ to a field generated by some potential $V_b$.
The minimization of the thermodynamic potential in the $b$-direction results in $b$ being a medium dependent quantity.
Since the potential $V_b$ is an additional contribution to the bag pressure, this 
prescription can be understood as a self-consistent way to generate a medium-dependent bag pressure. 
We therefore name this model the bag-quark-meson model (bQM), and the field $b$ is named the {\it bag} field.
We consider the bag field as a non-dynamical, auxiliary field.
This field is responsible for statistical confinement, in spirit 
similar to the Polyakov loop.

Taking into account the modification of the distribution functions, and the vacuum potential $V_b$, the 
thermodynamic potential of the model becomes
\be
\Omega = V_\sigma + V_b + \Omega_q~,
\label{eq:omthqtot}
\ee
where
\be
\Omega_q = \gamma_q \int\frac{d^3 p}{(2\pi)^3}
\left[T\log\left(1-n_q\right)+
T\log\left(1-\bar{n}_q\right)\right]~, 
\label{eq:omthq}
\ee
with $n_q$ given by (\ref{eq:ir}) and $\gamma_q = 2 \times N_f \times N_c = 12$ for two flavors.
The gap equations are
\be
\frac{\partial\Omega}{\partial \sigma} = -\lambda_\chi\chi_0^2\sigma+\lambda\sigma^3-\epsilon\chi_0^2+g_q s_q=0~,
\label{eq:gapqs}
\ee
and
\be
\frac{\partial\Omega}{\partial b} = \frac{\partial V_b}{\partial b}-\varpi_q=0~,
\label{eq:gapche}
\ee
where
\be
s_q = \gamma_q\int\frac{d^3 p}{(2\pi)^3}\frac{m_q}{E_q}(n_q + n_{\bar{q}} )~.
\label{eq:scal}
\ee
The term $\varpi_q$ is a boundary contribution of (\ref{eq:omthq})
\be
\varpi_q = \gamma_q\frac{b^2}{2\pi^2}\left[T\log(1-f_q)+T\log(1-\bar{f}_q)\right]_{\mathbf{p}^2 = b^2}~.
\label{eq:varpq}
\ee
The goal is to suppress quarks at low $T$ and/or $\mu_q$ with a large value for $b$, and to have lower values of $b$ at high $T$ and/or $\mu_q$.
Remarkably, this is accomplished through the minimization of the thermodynamic potential 
with respect to $b$: the thermal correction $\varpi_q$ to the gap equation for $b$ acts to {\it reduce} $b$.

\subsection{Parametrization of the bQM model}
\label{ssec:parbqm}

\begin{figure}[t]
\begin{center}
\includegraphics[clip,scale=0.38]{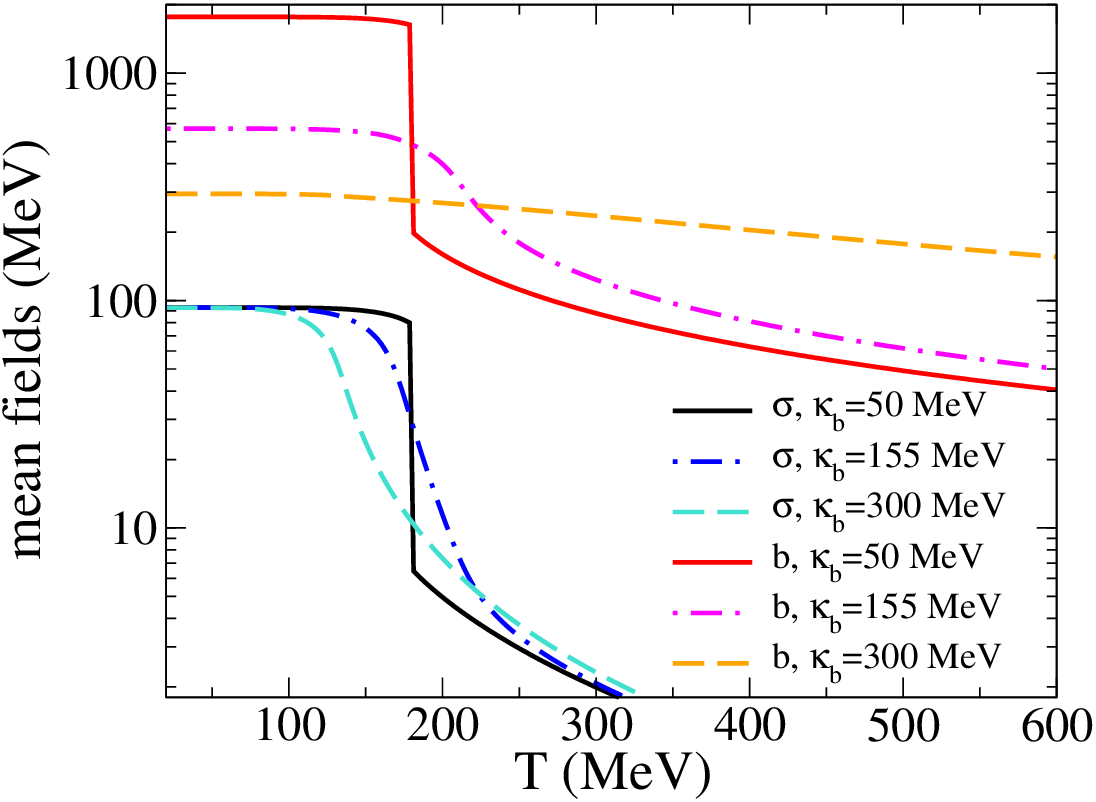}
\hspace{0.1cm}
\includegraphics[clip,scale=0.38]{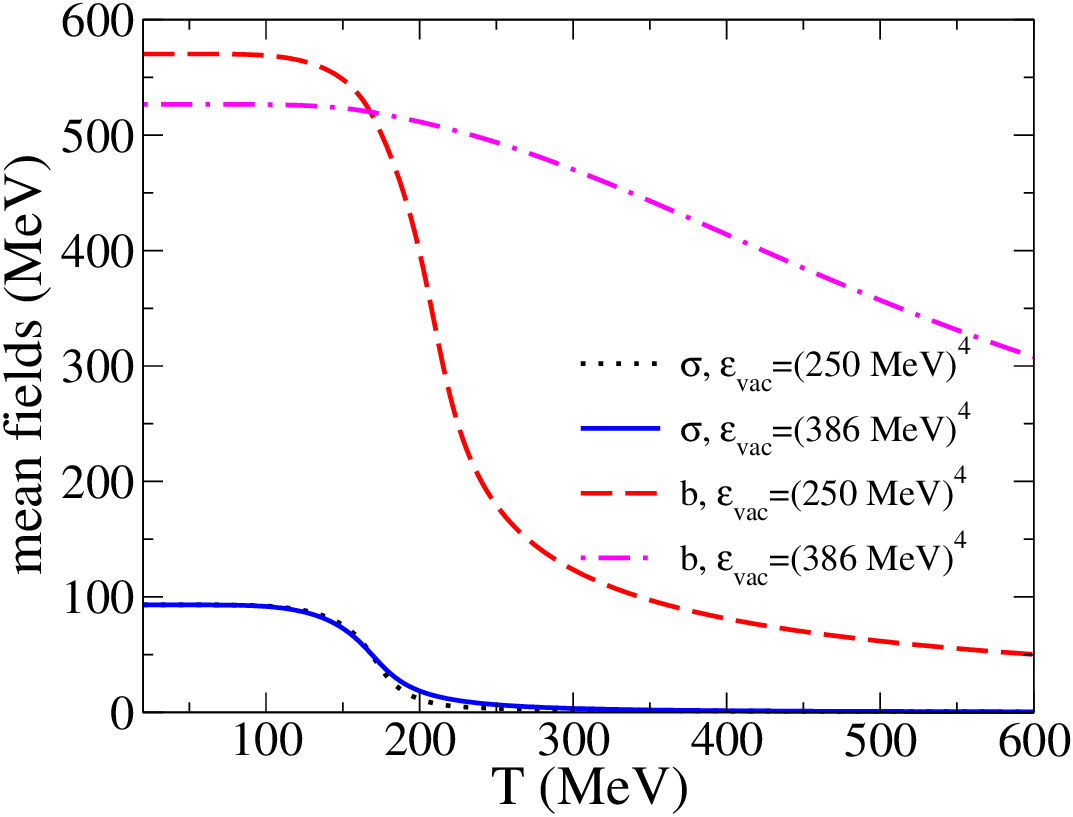}
\caption{The mean-fields $\sigma$ and $b$ in the bQM model as a function of $T$ at $\mu_q=0$. Left panel: results for different values of the parameter $\kappa_b$, with $\epsilon_{\rm vac} = (250 \, {\rm MeV})^4$.
Right panel: results for $\epsilon_{\rm vac}=(250 \, {\rm MeV})^4$, $\kappa_b = 155$ MeV and $\epsilon_{\rm vac}=(386 \, {\rm MeV})^4$, $\kappa_b = 400$ MeV.
In this case, the pseudo-critical temperature of the chiral transition is fixed to $T_c = 170$ MeV.}
\label{fig:bt2}
\end{center}
\end{figure}
Unlike the Polyakov loop potential, the potential $V_b$ cannot be constrained by symmetry, so
one should try different forms.
Below we consider a special case with a minimal number of terms that yields a finite value of $b$ in the vacuum.
Namely, we choose
\be
V_b = -\frac{\kappa_b^2}{2} b^2 +\frac{\lambda_b}{4} b^4~,
\label{eq:vb}
\ee
where the new parameters $\kappa_b$ and $\lambda_b$ must be determined. The non-trivial vacuum expectation value of this potential is $b_0 = \sqrt{\kappa_b^2/\lambda_b}$. One can further motivate this choice as follows. Since the $b$ field generates statistical confinement in our model, we associate it with the chromo-electric part of the gluon sector.
The essential reason for this is that finite $T$ lattice computations show that the chromo-electric part of the gluon condensate drops across the pseudo-critical temperature \cite{D'Elia:2002ck,Miller:2006hr,Colangelo:2013ila}.
Also, the chromo-electric sector is governed by the zeroth component 
of the gluon fields, like the Polyakov loop or the quark-antiquark potential, so it is essentially non-dynamical.
In that sense, such an identification can be considered natural.

Since the $b$-potential (\ref{eq:vb}) yields an additional contribution to the 
vacuum energy, we must ensure that the total vacuum energy remains correctly saturated.
We will consider a particular case where the vacuum energy created by the $\chi$-field and the $b$-field are equal in magnitude $V_{\chi_0} = V_{b_0}$,
so that a half of the total vacuum energy has its origin in the chromo-electric sector \cite{Lee:2009rz,Sasaki:2013xfa}, modeled here by $V_b$.
In this language, the dilaton $\chi$ naturally represents the chromo-magnetic component of the gluon condensate \cite{Lee:2009rz,Paeng:2011hy,Paeng:2013xya,Sasaki:2013xfa}, which survives the chiral transition.
We thus take
\be
-V_{b_0} = \frac{\kappa_b^4}{4\lambda_b} = \frac{\epsilon_{\rm vac}}{2}~,
\label{eq:vb0}
\ee
where the total vacuum energy is $\epsilon_{\rm vac} \simeq (193-386 \, {\rm MeV})^4$.
We fixed a value for $V_{b_0}$ (within the range for $\epsilon_{\rm vac}$) and then changed $\kappa_b$. 
We found that in general low values of $\epsilon_{\rm vac}$ and low values of $\kappa_b$ produce a first order 
chiral transition in the bQM model at $\mu_q=0$ and finite $T$, see the left panel Fig.~\ref{fig:bt2} for $\kappa_b = 50$ MeV.
Since lattice QCD shows that the chiral transition is a crossover this sets a lower bound on $\kappa_b$.
Increasing $\kappa_b$ while holding $\epsilon_{\rm vac}$ fixed leads to a decrease of $b_0$.
The result is that the chiral transition turns into a crossover with its onset shifted towards lower temperatures as shown on the left panel of Fig.~\ref{fig:bt2}.

As a second constraint, we 
choose
the pseudo-critical temperature $T_c$ as obtained in lattice QCD simulations.
For $N_f=2+1$ with physical quark masses the Wuppertal-Budapest collaboration obtained $T_c = 147(2)(3)$ MeV \cite{Borsanyi:2010bp}, while the HotQCD collaboration quoted $T_c = 154(9)$ MeV \cite{Bazavov:2011nk}.
Since in this paper we work with $N_f=2$ we will use a slightly larger value $T_c \simeq 170$ MeV \cite{Ejiri:2000bw,Maezawa:2007fd}.
This gives $\epsilon_{\rm vac} = (250 \, {\rm MeV})^4$ and $\kappa_b = 155$ MeV.
From (\ref{eq:vb0}) we find $\lambda_b=0.074$.
The resulting $\sigma$ and $b$ mean-fields obtained by solving (\ref{eq:gapqs}) and (\ref{eq:gapche}) at finite $T$ and $\mu_q=0$ are shown on Fig.~\ref{fig:bt2}, where $b_0 = 570.3$ MeV.

With the appropriate modification of 
the quark number density
\be
\rho_q = -\frac{\partial \Omega_q}{\partial \mu_q} = \gamma_q\int\frac{d^3 p}{(2\pi)^3}(n_q - \bar{n}_q)~.
\label{eq:num}
\ee
we find that the onset of quarks for $\epsilon_{\rm vac} = (250 \, {\rm MeV})^4$ in the bQM model
at $T=0$ is at $\mu_q \simeq 450$ MeV. 
This numerical value must be contrasted to the one in the QM model, where $\mu_q \simeq 250$ MeV.

It must be stressed that the obtained parameters $\kappa_b$ and $\lambda_b$ are not unique.
The uncertainty in the QCD vacuum energy provides a range for the parameters $\kappa_b$ and $\lambda_b$.
The fixed value of the vacuum energy $\epsilon_{\rm vac} = (250 \, {\rm MeV})^4$ is the {\it smallest} value which still satisfies the above-mentioned constraints.
Taking higher values of $\epsilon_{\rm vac}$, while keeping $T_c$ fixed we find that both $\kappa_b$ and $\lambda_b$ increase. Interestingly, the vacuum expectation value $b_0 = \sqrt{\kappa_b^2/\lambda_b}$ shows a reduction of less than $10\%$ in the range $\epsilon_{\rm vac} = (250-386 \, {\rm MeV})^4$.
The sensitivity to $\epsilon_{\rm vac}$ is reflected in the temperature at which the $b$-field experiences a rapid change. Higher values of $\epsilon_{\rm vac}$ represent a larger energy barrier for quark fluctuations so a delayed and more gradual change in the $b$-field is expected.
This is visible on the right panel of Fig.~\ref{fig:bt2} where the $b$ field is shown for $\epsilon_{\rm vac} = (250 \, {\rm MeV})^4$ and $\epsilon_{\rm vac} = (386 \, {\rm MeV})^4$.

\subsection{Similarities and differences between the bQM and the PQM model}
\label{ssec:sim}

\begin{figure}[t]		
\begin{center}
\includegraphics[clip,scale=0.4]{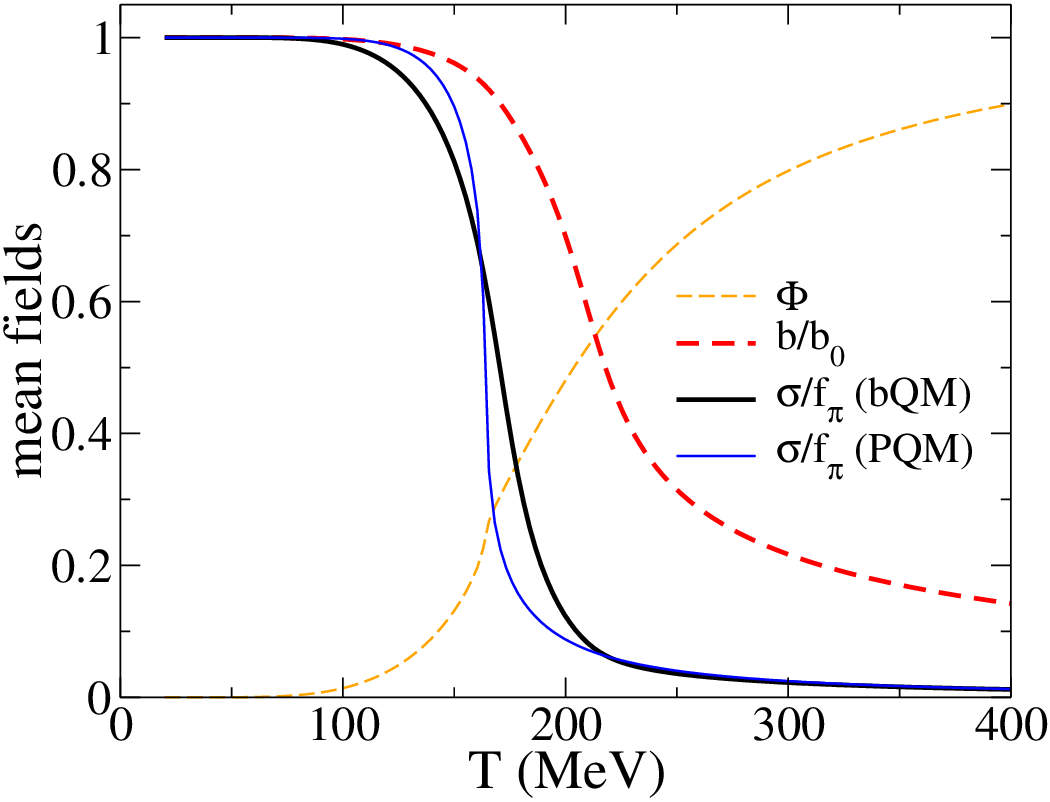}
\caption{A comparison between the normalized mean fields in the bQM and the PQM models as functions of $T$ at $\mu_q=0$.
The bQM results are presented for $\epsilon_{\rm vac} = (250 \, {\rm MeV})^4$.}
\label{fig:bt}
\end{center}
\end{figure}
Since a finite $b$ field reduces the strength of the quark thermal fluctuations, it is intuitively clear that it will act to increase the critical temperature associated with the chiral transition 
(see also left panel of Fig.~\ref{fig:bt2}).
In the chiral limit we can show this relation analytically by finding a zero of
the $\sigma^2$ coefficient in the Landau expansion of the thermodynamic potential (\ref{eq:omthqtot}) in powers of $\sigma^2$.
The result can be cast in the following parametric form
\be
-\lambda_\chi\chi_0^2 + \gamma_q g^2 \frac{T^2}{12}\mathcal{F}\left(\frac{b}{T}\right)=0~,
\label{eq:tc}
\ee 
where
\be
\mathcal{F}(x) = \frac{2}{\pi^2}(-3x^2+\pi^2+6x\log(1+e^x))+\frac{12}{\pi^2}{\rm Li}_2(-e^x)~,
\ee
and ${\rm Li}_2(x)$ is the polylogarithm of order 2.
In the limit $b/T\ll 1$ this can be simplified to
\be
T_c^{\rm bQM} \simeq \left(\frac{12 \lambda_\chi\chi_0^2}{\gamma_q g^2}+\frac{3}{\pi^2}b^2\right)^{1/2}~.
\ee
The second term under the square root is a correction to the usual QM model result for $T_c$.
It is instructive to compare this result to the one obtained in the Polyakov-quark-meson (PQM) model
\be
T_c^{\rm PQM} = \left(\frac{12 \lambda_\chi\chi_0^2}{\gamma_q g^2}+\frac{2}{\pi^2}\phi^2\right)^{1/2}~,
\ee
where $\phi$ is the background gauge field related to the Polyakov loop $\Phi$ through $\Phi = [1+2\cos(\phi/T)]/3$, see e.~g. \cite{Fukushima:2003fw}.

In the PQM model the Polyakov loop potential 
saturates the transverse (physical) gluon contribution to the thermodynamics.
On the other hand, in the bQM model the potential $V_b$ does not contain the transverse gluons.
This is the main difference between the PQM and the bQM models.
In the PQM model, deconfinement transition can be characterized by the Polyakov loop. Then, one can define the pseudo-critical temperatures governed by the peaks of the quark condensate and the Polyakov loop and interpret their approximate coincidence as obtained in lattice QCD simulations \cite{Borsanyi:2010bp}.
However, the bQM model in its present form cannot predict the deconfinement transition temperature, nor the coincidence of the two respective pseudo-critical temperatures, because the $b$ field is not a confinement-deconfinement order parameter.

To compare results for the bQM and the PQM models, we have calculated the $\sigma$ field and the Polyakov loop in a PQM model.
A concrete parametrization of the Polyakov loop potential is taken from \cite{Fukushima:2008wg}, while for the quark sector the same linear sigma model (\ref{eq:lsm}) is used. 
On Fig.~\ref{fig:bt} we compare our results where $\epsilon_{\rm vac}$  in the bQM model was set to $\epsilon_{\rm vac} = (250 \, {\rm MeV})^4$.
We find that the $\sigma$ field in both models are closely matched, with peaks in $d\sigma/dT$ at $163$ MeV and $170$ MeV (the fitted value), respectively.
Obviously, a more realistic consideration should include explicitly the bosonic excitations.

\section{A combined quark-meson-nucleon model}
\label{sec:qmn}

In this section we construct a combined quark-meson-nucleon (QMN) model.
Our guiding requirement is to exclude quarks at low density, and to exclude nucleons at high density.
In the previous Section we have introduced the concept of statistical confinement of quarks, and now we propose a similar but opposite
modification of the distribution functions for nucleons
\be
n_{N_\pm} = \theta(\alpha^2 b^2 - \mathbf{p}^2)f_{N_\pm}~, \qquad \bar{n}_{N_\pm} = \theta(\alpha^2 b^2 - \mathbf{p}^2)\bar{f}_{N_\pm}
\label{eq:uv}
\ee
With this ansatz nucleon fluctuations with momenta $\mathbf{p}^2>(\alpha b)^2$ are suppressed, where $\alpha$ is a new parameter of the model.
A similar form has been used in \cite{Berges:1998ha}.
In Section \ref{ssec:param} we will determine
the possible range for the parameter $\alpha$.  

\subsection{Model setup}

The thermodynamic potential of the QMN model is obtained as the following sum
\be
\Omega = V_\sigma+V_\omega+V_\chi+V_b + \sum_{X=N_\pm, q}\Omega_X~,
\label{eq:tot}
\ee
with $\Omega_{N_\pm}$ defined as in (\ref{eq:omthn}) with the appropriate modification of the distribution functions $f_{N_\pm} \to n_{N_\pm}$ according to Eq.~$(\ref{eq:uv})$, while $\Omega_q$ is defined in (\ref{eq:omthq}).

The gap equations are obtained by minimizing the thermodynamical potential (\ref{eq:tot}).
The $\omega$ and $\chi$ gap equations remain unchanged by the inclusion of quarks, being given by Eqs.~(\ref{eq:gapom}) and (\ref{eq:gapch}), respectively, and the proper replacement $f_{N_\pm} \to n_{N_\pm}$ according to Eq.~(\ref{eq:uv}).
Thee gap equations for $\sigma$ and $b$ are modified as follows
\be
\frac{\partial\Omega}{\partial \sigma} = -\lambda_\chi\chi^2\sigma+\lambda\sigma^3-\epsilon\chi^2+
\sum_{X=N_\pm,q}\frac{\partial m_X}{\partial\sigma}s_X=0~,
\label{eq:gaps2}
\ee
\be
\frac{\partial\Omega}{\partial b} = -\kappa_b^2 b+\lambda_b b^3+
\alpha\sum_{X=N_\pm}\varpi_X-\varpi_q=0~,
\label{eq:gapb}
\ee
where
\be
\varpi_{N_\pm} = \gamma_N\frac{(\alpha b)^2}{2\pi^2}\left[T\log(1-f_{N_\pm})+T\log(1-\bar{f}_{N_\pm})\right]_{\mathbf{p}^2 = (\alpha b)^2}~,
\ee
and $\varpi_q$ is given in (\ref{eq:varpq}).

The pressure is calculated by evaluating the thermodynamic potential at its minimum $p = -\Omega + \Omega_0$, normalized with the constant $\Omega_0$ in such a way that the physical vacuum has zero pressure.
The total baryon number density is
$$\rho_B = -\frac{\partial \Omega}{\partial\mu_B} = \rho_{N_+} +\rho_{N_-} +\frac{1}{3}\rho_q~,$$
where we used $\mu_q = \mu_B/3$.
The nucleon and quark particle fractions are defined as
\be
Y_{N_\pm} = \frac{\rho_{N_\pm}}{\rho_B}~, \qquad
Y_q = \frac{1}{3}\frac{\rho_q}{\rho_B}~.
\label{eq:frac}
\ee

For the transition from nucleon to quark degrees of freedom the gap equation for the $b$-field, Eq.~(\ref{eq:gapb}), is crucial.
Note that, as a consequence of the Leibniz rule, the nucleon and quark contributions in (\ref{eq:gapb}) have opposite signs.
In the low density phase nucleons will favor finite $b$ and as a consequence quarks are suppressed.
The appearance of quarks at high densities acts to {\it reduce} $b$ and therefore exclude nucleons.
Therefore, Eq.~(\ref{eq:gapb}) controls the relative abundance of nucleons and quarks.
In order to better illustrate this point we have solved a simplified model with only the $b$ field in the Appendix.

\subsection{Parametrization}
\label{ssec:param}

\begin{table}[htb]
\begin{center}
\begin{tabular}{|c|c|c|c|c|c|c|c|c|c|c|c|c|}
\hline
$m_\sigma$ (MeV) & $m_\chi$ (MeV) & $B$ (MeV)$^4$ & $g_1$ & $g_2$ & $g_\chi$ & $g_\omega$ & $g_q$ & $\lambda$ & $\lambda_\chi$ & $\kappa_b$ (MeV) & $\lambda_b$\\
\hline
\hline
370.63 & 1700 & 297.30 & 13.0 & 6.97 & 4.39 & 6.79 & 3.22 & 6.84 & 3.71 & 155 & 0.074\\
\hline
\end{tabular}
\end{center}
\caption{The parameters of the quark-meson-nucleon model.}
\label{tab:par}
\end{table}
The parameters of the vacuum potential (as well as the effective masses $m_X$ and chemical potentials $\mu_X$) are defined in Sec.~\ref{sec:nuc} and Sec.~\ref{sec:qu}, and collected in Table \ref{tab:par}. 
In particular, we will use $V_b$ with $\epsilon_{\rm vac} = (250 \, {\rm MeV})^4$ as discussed in Sec.~\ref{ssec:parbqm}.
The remaining parameter $\alpha$ is chosen so that the effective UV cutoff $\alpha b_0$ for the nucleon distribution functions does not spoil the nuclear matter ground state.
This sets the minimal value for $\alpha b_0 \gsim 300$ MeV.
Assume now that at some $\mu_B$ there is a transition to quark matter.
From the following consideration this will lead to a useful estimate on the upper bound on $\alpha$.
Consider Eq.~(\ref{eq:gapb}) in the limit of large $\mu_B$.
At $T\to 0$ the boundary terms read
\be
\varpi_{N_\pm} \to -\gamma_N\frac{(\alpha b)^2}{2\pi^2}(\mu_N - E_{N_\pm})\theta(\mu_N - E_{N_\pm})~,
\label{eq:surf1}
\ee
\be
\varpi_q \to -\gamma_q\frac{b^2}{2\pi^2}(\mu_q - E_q)\theta(\mu_q - E_q)~.
\label{eq:surf2}
\ee
We assume $\sigma, \, \omega\to 0$ in this limit.
In addition, we consider a case where $\mu_B$ is large enough that we can also ignore the chirally invariant nucleon mass.
Using (\ref{eq:surf1}), Eq.~(\ref{eq:gapche}) simplifies to
\be
\frac{\partial\Omega}{\partial b} = -\kappa_b^2 b+\lambda_b b^3+
\frac{b^2}{2\pi^2}
\left(-2\alpha^3\gamma_N+\frac{\gamma_q}{3}\right)\mu_B=0~.
\label{eq:gapb2}
\ee
Since $\gamma_N = \gamma_q/3$, the vanishing 
bracket defines $\alpha_{\rm max} = 2^{-1/3}$.
For $b_0 = 570.3$ MeV we have $\alpha_{\rm max} b_0 = 452.6$ MeV.
In the following results we consider two values of the $\alpha$ parameter: $\alpha b_0 =300$ MeV and $\alpha b_0 = 440$ MeV, which is close to the benchmark value $\alpha_{\rm max} b_0 = 452.6$ MeV.
We also calculate the transition points for several $\alpha$'s in between.

\section{Results}
\label{sec:res}

\begin{figure}[t]
\begin{center}
\includegraphics[clip,scale=0.45]{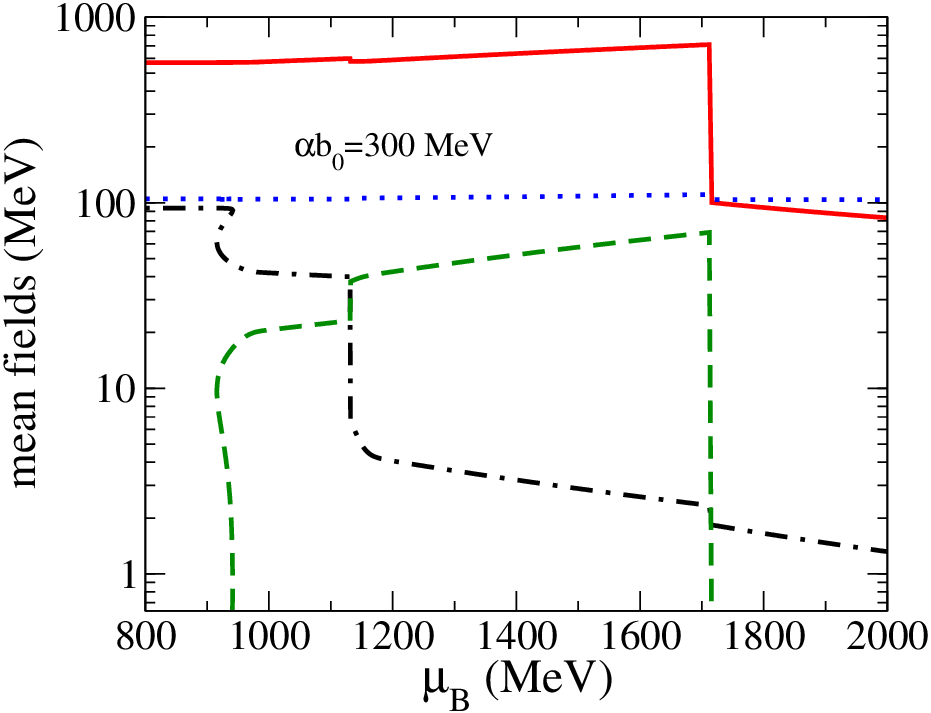}
\hspace{0.1cm}
\includegraphics[clip,scale=0.45]{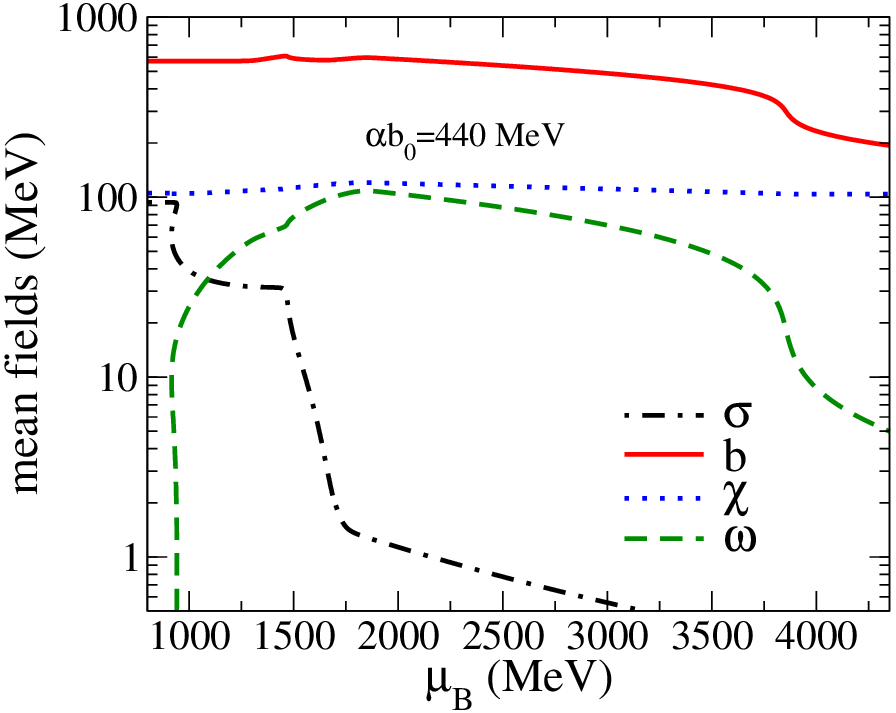}
\caption{The mean fields for the QMN model
for two different values of the $\alpha$ parameter indicated in the figure.
As $\mu_B$ is increased there is a clear imprint of the liquid-gas and the chiral phase transitions on the $\sigma$ field shown by the dash-dot black line.
The drop in the $b$ field at high $\mu_B$ is associated with the deconfinement transition.}
\label{fig:mfqn}
\end{center}
\end{figure}
Without the modification of the nuclear distribution functions the parity doublet model with dilaton introduced in Sec.~\ref{sec:nuc} has a weak first order chiral transition at $\mu_B \simeq 2110$ MeV for the parameters used here.
This should be contrasted to the value of $\mu_B\simeq 1725$ MeV found for the same parameter set \cite{Zschiesche:2006zj,Sasaki:2010bp} but using a model without the dilaton.
Such a delay in the chiral transition takes place because the dilaton field is essentially coupled to the density through the $\omega$ field, see Eq.~(\ref{eq:vom}).
In the case where the distribution functions are modified 
we must deal with the uncertainty in the parameter $\alpha$ and $\epsilon_{\rm vac}$.
We show explicit calculations for the case $\epsilon_{\rm vac} = (250 \, {\rm MeV})^4$ and for two limiting values of $\alpha$.
Finally, we will discuss sensitivity to the uncertainty in $\epsilon_{\rm vac}$.

With $\alpha b_0 = 300$ MeV the onset of chiral-symmetric phase occurs at a lower $\mu_B$.
This can be seen on Fig.~\ref{fig:mfqn} showing the solutions of the gap equations in the QMN model.
This is easily understood: in this model it is the nucleons that restore the chiral symmetry - quarks appear only at higher $\mu_B$, see Fig.~\ref{fig:densq}.
The cutoff in the nucleon momenta limits their density, and the $\omega$ field reaches a plateau after $\mu_B \simeq 1000$ MeV.
Therefore, the shift of $\mu_B$ due to the $\omega$ field is diminished and, as a consequence, the chiral transition happens at lower $\mu_B$ than in the model without a cutoff.
At $\alpha b_0 = 300$ MeV
the chiral transition is first order and occurs at $\mu_B = 1135$ MeV.
After this point, the $\sigma$ field drops almost to zero and the parity-doublet partners become degenerate and equally populated.
In the case $\alpha b_0 = 440$ MeV, the chiral phase transition is a crossover.
From the peak in $d\sigma/d \mu_B$ we extract $\mu_B = 1473$ MeV.
At that point the mass splitting of the parity partners is around $18\%$ of its chirally invariant contribution in the vacuum given by $g_\chi \chi_0=790$ MeV.

\begin{figure}[t]
\begin{center}
\includegraphics[clip,scale=0.45]{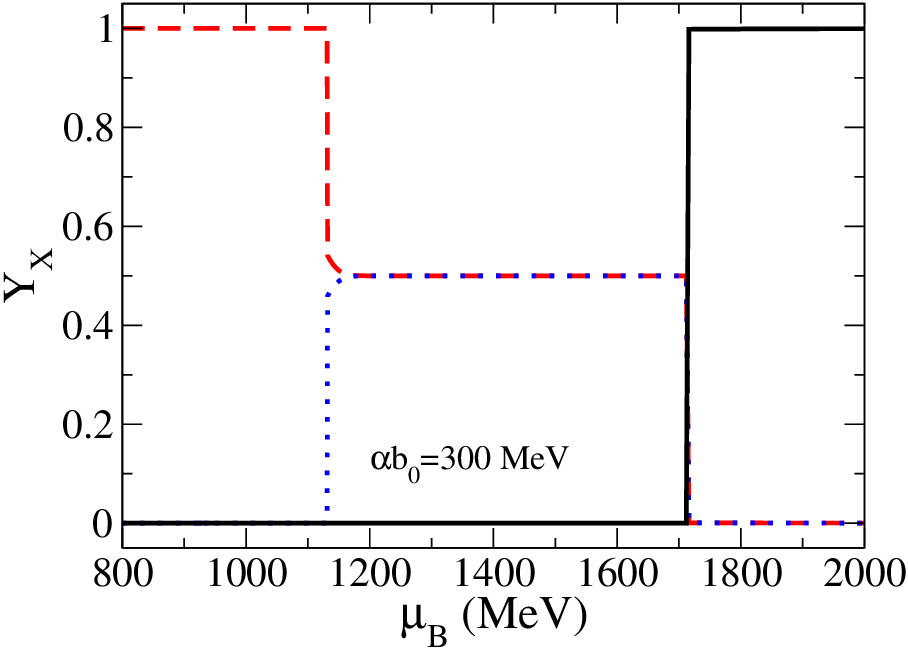}
\hspace{0.1cm}
\includegraphics[clip,scale=0.45]{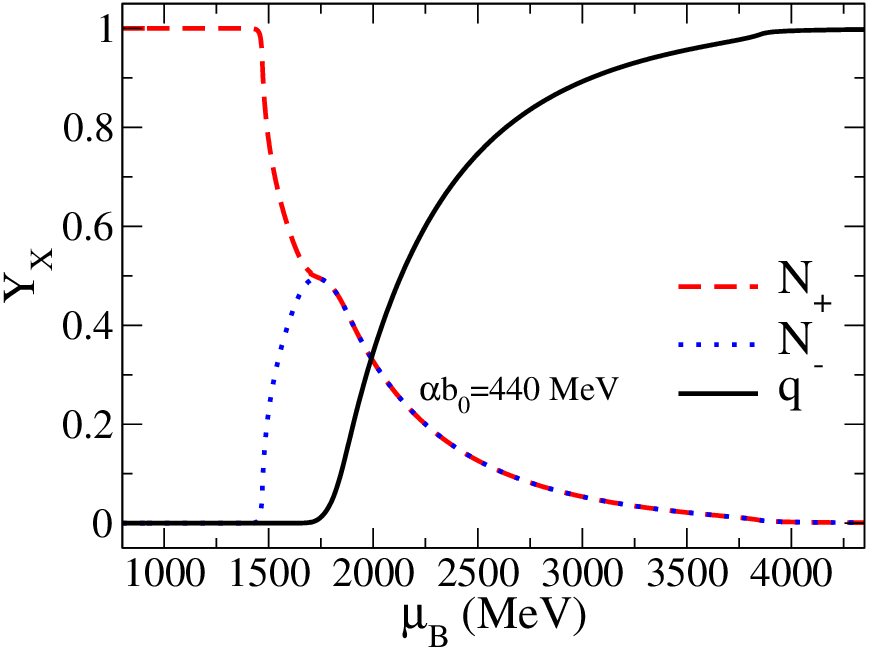}
\caption{Quark and nucleon particle fractions for the QMN model for two different values of the $\alpha$ parameter as indicated in the figure.}
\label{fig:frac}
\end{center}
\end{figure}
At some higher $\mu_B$ the $b$-field decreases, as can be seen on Fig.~\ref{fig:mfqn}.
The reduction of the $b$-field suppresses nucleons simultaneously enhancing quark fluctuations, according to Eqs.~(\ref{eq:uv}) and (\ref{eq:ir}), respectively.
Therefore, this marks the deconfinement transition in the QMN model.
This is demonstrated on Fig.~\ref{fig:frac} where we plot the nucleon and quark particle fractions defined in Eq.~(\ref{eq:frac}).

\begin{figure}[t]
\begin{center}
\includegraphics[clip,scale=0.4]{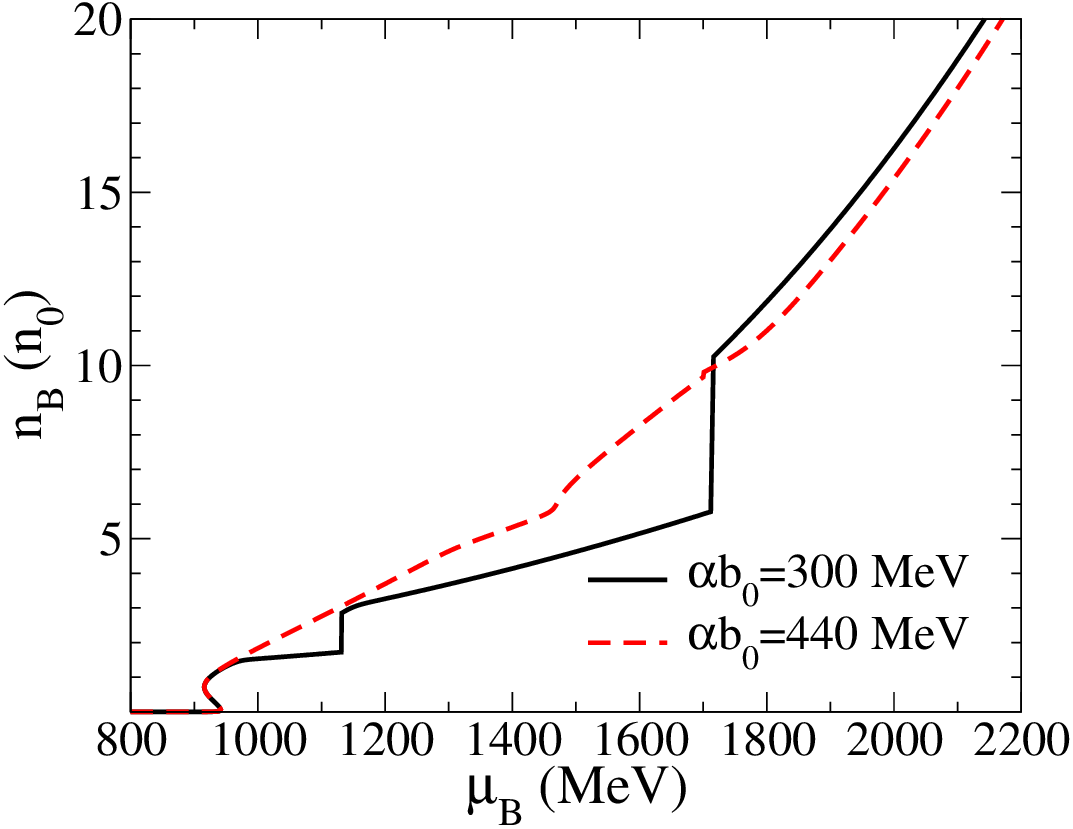}
\caption{The baryon number density $n_B$ (in units of saturation density $n_0$) as a function of the baryon chemical potential $\mu_B$ in the QMN model and for two different values of the $\alpha$ parameter indicated in the figure.}
\label{fig:densb}
\end{center}
\end{figure}

The strength of the transitions strongly depends on the value of $\alpha$.
For $\alpha b_0 = 300$ MeV the deconfinement transition is accompanied by a jump in the $b$ field - consequently the baryon density has a jump, so it is rightful to consider this as a first order phase transition.
On the other hand, for $\alpha b_0 = 440$ MeV, the $b$-field reduces gradually so that the deconfinement transition is in fact a crossover, allowing for a wide region of a mixed phase of nucleons and quarks, see Fig.~\ref{fig:frac}.
In this case we conventionally mark the point of the deconfinement transition with $\mu_B$ at which $Y_q = Y_{N_\pm}$ holds. The model predicts that the deconfinement transition happens at $\mu_B^{\rm d} = 1716$ MeV for the case $\alpha b_0 = 300$ MeV, while it occurs at $\mu_B^{\rm d} = 2130$ MeV, for $\alpha b_0 = 440$ MeV.
The difference between these two cases is seen for example in the behavior of
particle fractions on Fig.~\ref{fig:frac}.
  
\begin{figure}[t]
\begin{center}
\includegraphics[clip,scale=0.4]{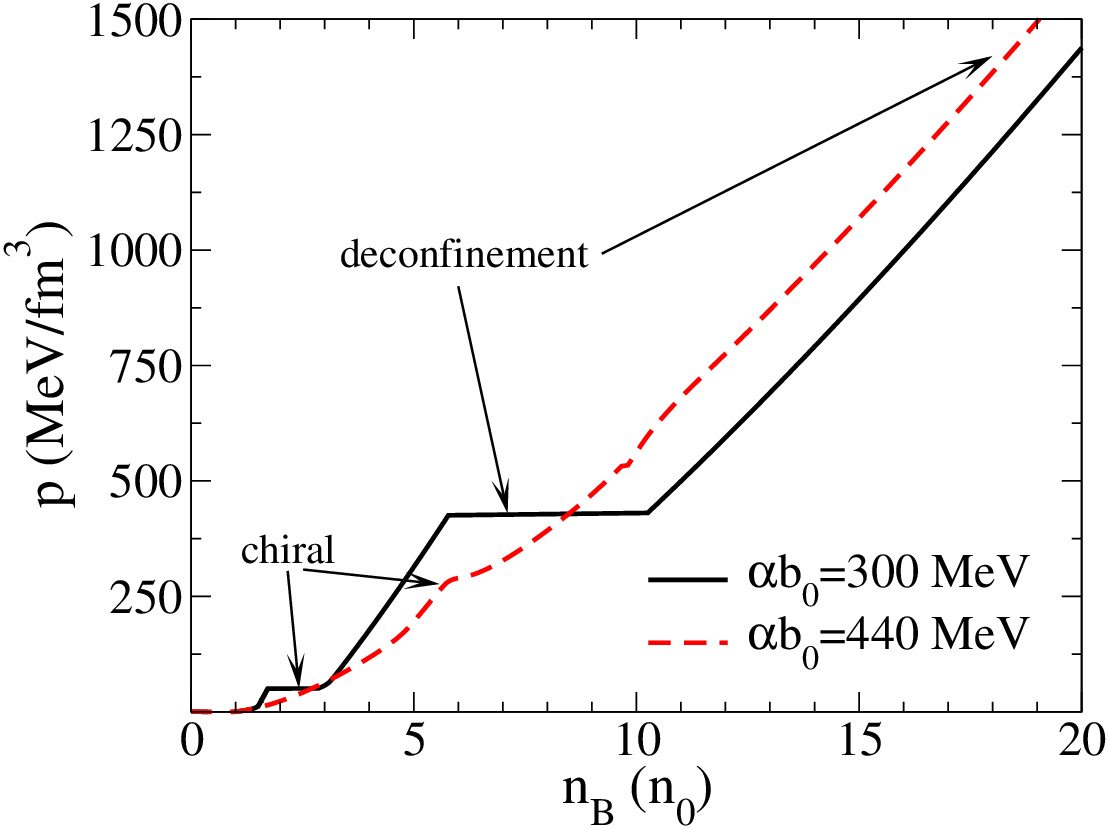}
\caption{The equation of state for the QMN model for two different values of the $\alpha$ parameter indicated in the figure.
The locations of the chiral and the deconfinement transition are approximately marked. In the case of $\alpha b_0 = 440$ MeV, the arrow marks the point where $Y_{N_+} + Y_{N_-} = Y_q$.}
\label{fig:eosqcd}
\end{center}
\end{figure}  
The chiral and the deconfinement transition are reflected in the behavior of the density $n_B$ as a function of $\mu_B$, plotted on Fig.~\ref{fig:densb}, and the equation of state, plotted on Fig.~\ref{fig:eosqcd} in the $p-n_B$ plane.
While for both values of $\alpha$ the chiral transition is visible as a small density change, 
for $\alpha b_0=300$ MeV the deconfinement transition has a pronounced density jump in the region $n_B \simeq 6-10 n_0$.
On the other hand, with $\alpha b_0 = 440$ MeV the deconfinement transition is continuous and there is no clear imprint on the equation of state.

\begin{table}[htb]
\begin{center}
\begin{tabular}{|c|c|c|c|c|c|c|}
\hline
$\alpha b_0$ (MeV) & $\mu_B^{\rm ch}$ (MeV) & $n_B^{\rm ch}$ ($n_0$) & order & $\mu_B^{\rm d}$ (MeV) & $n_B^{\rm d}$ ($n_0$) & order\\
\hline
\hline
300 & 1132 & 1.7 & 1st order & 1716 & 5.8 & 1st order\\
\hline
350 & 1220 & 2.8 & 1st order & 1851 & 9.0  & 1st order\\
\hline
400 & 1348 & 4.4 & crossover & 1931 & 11.7 & 1st order\\
\hline
440 & 1473 & 6.1 & crossover & 2130 & 18.1 & crossover\\
\hline
\end{tabular}
\end{center}
\caption{The baryon chemical potential $\mu_B$ and the baryon number density $n_B$ at the onset of chiral and the deconfinement phase transitions for several values of $\alpha$ in the QMN model.
We also denote the order of the chiral and the deconfinement transition for both cases.
In the case of a first order transition, the transition $\mu_B$ is defined as a lower value of the number density jump.
In the case of a crossover, $\mu_B^{\rm ch}$ is defined by the peak in $d\sigma / d\mu_B$, while $\mu_B^{\rm d}$ is defined as the point where $Y_{N_+} + Y_{N_-}=Y_q$.
The values of $n_B$ are given in units of the saturation density $n_0 = 0.16$ fm$^{-3}$.}
\label{tab:params}
\end{table}
In Table \ref{tab:params} we collect the numerical values for the chemical potential and the total baryon density at the onset of the chiral and the deconfinement phase transition for several values of $\alpha$.
With increasing $\alpha$, both the chiral and the deconfinement transition are shifted to higher $\mu_B$.
The onset of the chiral phase transition is limited by the value $\mu_B \simeq 2100$ MeV, which is the result for any $\alpha>\alpha_{\rm max}$.
On the other hand, the deconfinement phase transition will be pushed to $\mu_B\to \infty$ as $\alpha\simeq\alpha_{\rm max}$ is approached.
Beyond $\alpha_{\rm max}$ there is no deconfinement transition.
To summarize, we have found that the chiral and the deconfinement transitions in the model do not coincide for any choice of $\alpha$.

Lattice QCD results predict that the chiral and the confinement transitions approximately coincide along the finite $T$ axes of the QCD phase diagram.
We have performed calculations at two non-zero temperatures, $T=50$ MeV and $T=100$ MeV to test whether such a tendency may be observed within the QMN model.
At $T=0$ MeV the separation between the two transitions is roughly $\Delta \mu_B \sim 580-660$ MeV (see Table \ref{tab:params}), within the explored $\alpha$ range.
With $T=50$ MeV we find $\Delta \mu_B \sim 560-880$ while at $T=100$ MeV we find $\Delta \mu_B \sim 420-850$ MeV, i.~e. there is no clear trend in the critical parameter.

Taking into account the uncertainty in $\epsilon_{\rm vac} = (250-386 \, {\rm MeV})^4$ does not lead to a significant modification in the above results.
We have explicitly checked that for $\epsilon_{\rm vac} = (386 \, {\rm MeV})^4$ the $b$ field changes more gradually with increasing $\mu_B$ (similar to the finite $T$ results presented in Fig.~\ref{fig:bt2}) thereby softening the chiral and the deconfinement transition.

\section{Conclusions}
\label{sec:conc}

The properties of matter at large baryon densities are almost exclusively considered either in purely nucleonic models, or within purely quark models, with rare attempts to a unified description \cite{Berges:1998ha,Meyer:1999bx,Lawley:2006ps,Dexheimer:2009hi,Steinheimer:2011ea}.
A crucial ingredient for a unified description should be an effective mechanism to exclude quarks in the dilute hadronic matter, and to exclude nucleons at asymptotically high densities.
While the former has been accomplished by coupling quarks to the Polyakov loop, it is applicable only at finite temperatures and low densities.
In this work we have used a simplistic modification of the distribution functions of both the nucleons and the quarks that provides a mechanism to exclude quarks at low density and nucleons at high density.

We have considered nuclear matter in a parity doublet model with mirror assignment coupled to the dilaton field.
We have argued that the chiral potential which is fitted to the nuclear ground state properties is shallow (i.~e. the $\sigma$ meson is light).
When the same potential is used for quark matter it yields an early onset of quarks.
This is especially acute at zero temperature, where we find that with such a chiral potential the onset of quarks happens at baryon chemical potential below the nucleon mass. 
Chiral models of nuclear
matter usually favor low $m_\sigma$, see e.~g. \cite{Boguta:1982wr,Ellis:1992ey,Mishustin:1993ub,Zschiesche:2006zj}, so the problem of the shallow potential seems to be somewhat general. 
However, we stress that it is a problem only if we choose to couple the quarks to the same $\sigma$ field.

We have proposed a possible solution of this problem by introducing the concept of statistical confinement for quarks.
The quark distribution functions were modified in such a way that the quarks are suppressed below some particular momentum.
We associate this minimum momentum with an auxiliary scalar field, named the bag field.
This field brings a new contribution to the 
total vacuum potential.
We have found that the minimization of the thermodynamic potential in the $b$ direction provides a thermodynamically consistent framework.
Moreover, the $b$ field is finite in the vacuum and is reduced as temperature or density is increased.
At finite temperatures the mechanism of suppression of
quarks is similar to the effect of the Polyakov loop.
We have proposed a phenomenological form for the bag potential and fitted its parameters to the QCD vacuum energy $\epsilon_{\rm vac}$ and to $T_c$ at $\mu_B=0$ known from the lattice simulations.
Within this scheme it was possible to solve the problem of a shallow potential.

Further on, we have generalized the nucleonic distribution functions, by restricting their ultraviolet momentum space up to a value $\alpha b$ where $\alpha$ is an additional parameter in the model.
We have used the same bag potential to construct a combined quark-meson-nucleon model.
The vital feature of this model is that 
both the quarks and the nucleons are coupled to the same bosonic fields.
Then the fate of the QCD phase transitions was investigated at finite densities.

We explored a range of values for the parameter $\alpha$ and found that an increase in $\alpha$ delays the onset of the chiral and the deconfinement transitions.
The requirements that the nuclear ground state is not affected by quarks, and that the deconfinement transition happens at some density, restrict $\alpha$ to a finite range.
Taking into account the uncertainty in $\epsilon_{\rm vac}$ within this range of $\alpha$ we find that the chiral transition occurs at $\mu_B^{\rm ch} \simeq 1100 - 2100$ MeV.
For the deconfinement transition we predict a lower bound $\mu_B^{\rm d} \gsim 1700$ MeV.
In terms of the density the corresponding values are
$n_B^{\rm ch} \simeq 1.5-15.5 n_0$ and $n_B^{\rm d}\gsim 5n_0$.
For the lowest value $\epsilon_{\rm vac} = (250 \, {\rm MeV})^4$ order of both phase transitions depends on the value of particular $\alpha$.
With low values of $\alpha$ both transitions are first order, while for the higher values of $\alpha$ they are smooth crossovers.
Especially, the deconfinement transition proceeds in a broad crossover mixed phase where nucleons and quarks coexist.
These features persist as we vary $\epsilon_{\rm vac}$.

In this model the chiral and the deconfinement transition are always separated. The chiral transition is driven by the nucleonic fluctuations for any $\alpha$.
It is suggestive to consider this result in the light of the calculation of the Wilson and the Polyakov loop on the lattice with Dirac zero modes artificially removed. In Refs.~\cite{Gongyo:2012vx,Doi:2014zea} it was found that in this chirally ``unbroken'' phase the Wilson loop still displays an area law, and that the Polyakov loop is almost zero.
The separation of the transitions might be considered as a manifestation of the quarkyonic phase \cite{Hidaka:2008yy,McLerran:2008ua}, where chiral symmetry is restored in the nucleonic phase\footnote{It is important to stress this does not mean that the nucleons are massless, but that the parity partners are degenerate.}.

The separation of the chiral and the deconfinement transition persists also at non-zero $T$, where we checked two values $T=50$ MeV and $T=100$ MeV.
However, we do not expect the QMN model itself to be valid at such high $T$ where gluons and lightest hadrons become important.
Since in the QMN model we cannot calculate the Polyakov loop, the deconfinement transition in this case must be characterized in a different way. 
For these reasons, the fit of $V_b$ to lattice $T_c$, performed here only in the very simple bQM model, carries systematic uncertainties.
In general, the comparison to lattice QCD should be revised within a more complete approach before addressing the full $T-\mu_B$ phase diagram.

We must emphasize that the bag field should not be considered as an order parameter in this model, since it is not connected to any of the fundamental QCD symmetries.
In particular, it must not be considered as an order parameter for the deconfinement transition, even though it does play a key role in establishing it.
The value of $\mu_B$ where the transition occurs is for some values of the $\alpha$ parameter accompanied by a finite jump in the total baryon density.
In these cases, the density contrast between two coexisting phases can be used to characterize the deconfinement phase transition.
We can draw an immediate analogy to the liquid-gas phase transition which does not have an order parameter related to a symmetry, but is also characterized by a jump in the density from the liquid to the gas phase.

The approach to the deconfinement transition presented in this work
is based on the phenomenological requirements
and not rigorously grounded in QCD.
Due to this, systematic uncertainties are expected, especially concerning the freedom in choice of the bag potential and the explicit form for the modification of the distribution functions. 
It would be very interesting to obtain a first-principle information about the finite density distribution functions from Dyson-Schwinger studies, see e.~g. \cite{Klahn:2009mb}.
This microscopic input could then be used to construct more realistic effective models of nuclear and quark matter.

\subsection*{Acknowledgments}

The visits of S.~B. to the Frankfurt Institute of Advanced Studies were supported by
the mobility programme of the University of Zagreb and by the COST Action MP1304 “NewCompStar” within the STSM
programme.
S.~B. acknowledges partial support by the Croatian Science Foundation under Project No.~8799.
The work of I.~M. and C.~S. was
partly supported by the Hessian LOEWE initiative through the
Helmholtz International Center for FAIR (HIC for FAIR).
I.~M. acknowledges a partial support from grant NS-932.2014.2 (Russia). C.~S. acknowledges a partial support by the Polish
Science Foundation (NCN) under Maestro grant DEC-2013/10/A/ST2/00106.

\section{appendix}

This appendix is devoted to solving a simplified version of the quark-meson-nucleon model, where we focus only on the role of the bag field $b$.
The masses of nucleons and quarks are held fixed to $1000$ MeV and $0$ MeV, respectively.
With all the other mean-fields except the $b$ field discarded, the thermodynamic potential takes the following form
\be
\begin{split}
\Omega = V_b + \sum_{X=q,N_\pm} \gamma_X \int \frac{d^3 p}{(2\pi)^3}
\Big[ T &\log(1- n_X)\\
&+T\log(1- \bar{n}_X)\Big]~.
\end{split}
\label{eq:pots}
\ee
For the potential $V_b$ we use the same Eq.~(\ref{eq:vb}) with the parameters $\kappa_b = 155$ MeV, $\lambda_b = 0.074$.
For the parameter $\alpha$ we take the value $\alpha b_0 = 250$ MeV.
The partial pressures of quarks ($p_q$) and nucleons ($p_N$) are calculated as
$$p_q = -(V_b-V_{b_0}) - \Omega_q~,$$
$$p_N = -\Omega_{N_+} - \Omega_{N_-}~,$$
where the pressure contribution arising from $V_b$ is naturally assigned to quarks.

\begin{figure}[t]
\begin{center}
\includegraphics[clip,scale=0.4]{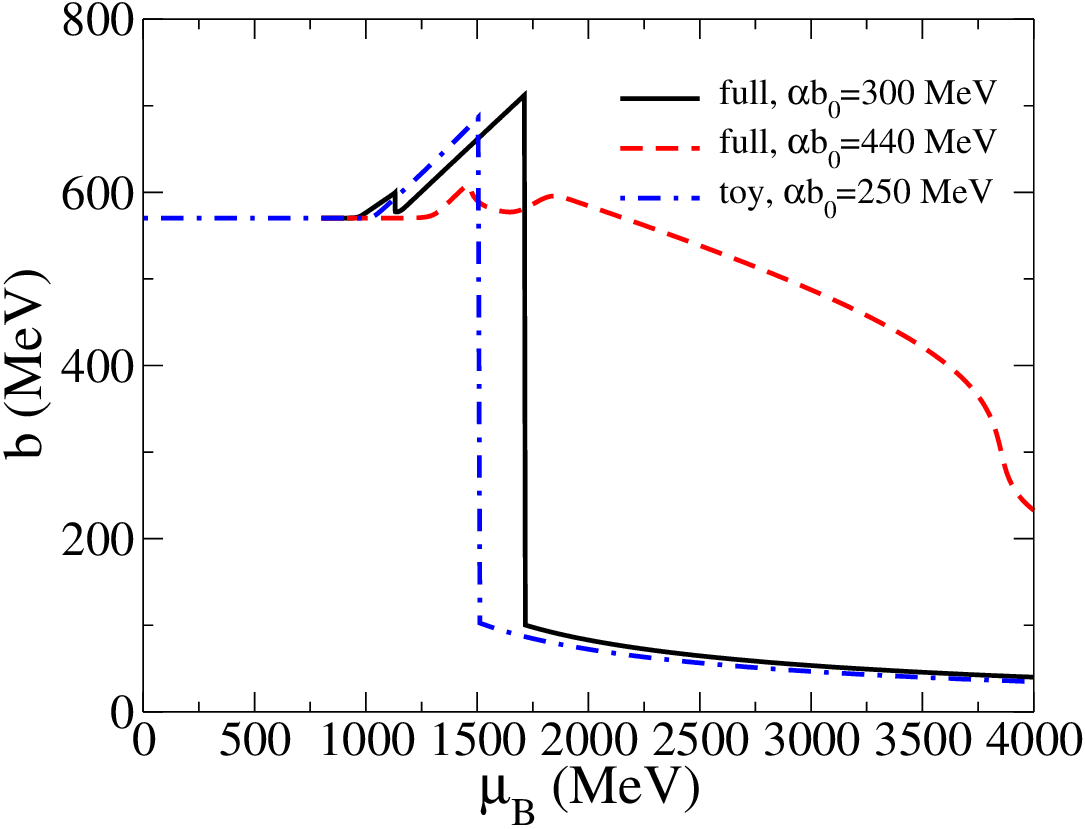}
\caption{The bag field as a function of $\mu_B$ in the simplified quark-meson-nucleon model, denoted with blue, dash-dotted line.
For comparison we also plot the bag field from the full quark-meson-nucleon model for two different values of $\alpha$ calculated in Sec.~\ref{sec:res}.}
\label{fig:bmju}
\end{center}
\end{figure}
From the minimization of the thermodynamic potential (\ref{eq:pots}) we obtain $b$ as a function of $\mu_B$ shown on Fig.~\ref{fig:bmju}.
The $b$ field follows its vacuum value up to $\mu_B = 1000$ MeV, then it begins to increase.
The increase is due to the fact that nucleons favor a finite value of $\mu_B$ according to Eq.~(\ref{eq:gapb}).
We expect the onset of quark degrees of freedom around $\mu_B \simeq 3 b_0 \simeq 1700$ MeV.
An explicit calculation gives $\mu_B=1515$ MeV.
This point marks the sudden drop of the $b$-field as seen in Fig.~\ref{fig:bmju}.
The qualitatively similar characteristics are found in Sec.~\ref{sec:res} in the complete model for $\alpha b_0 = 300$ MeV, also shown in Fig.~\ref{fig:bmju}.
However, for higher $\alpha b_0 = 440$ MeV the transition becomes gradual and shifts to higher $\mu_B$.

\begin{figure}[t]
\begin{center}
\includegraphics[clip,scale=0.4]{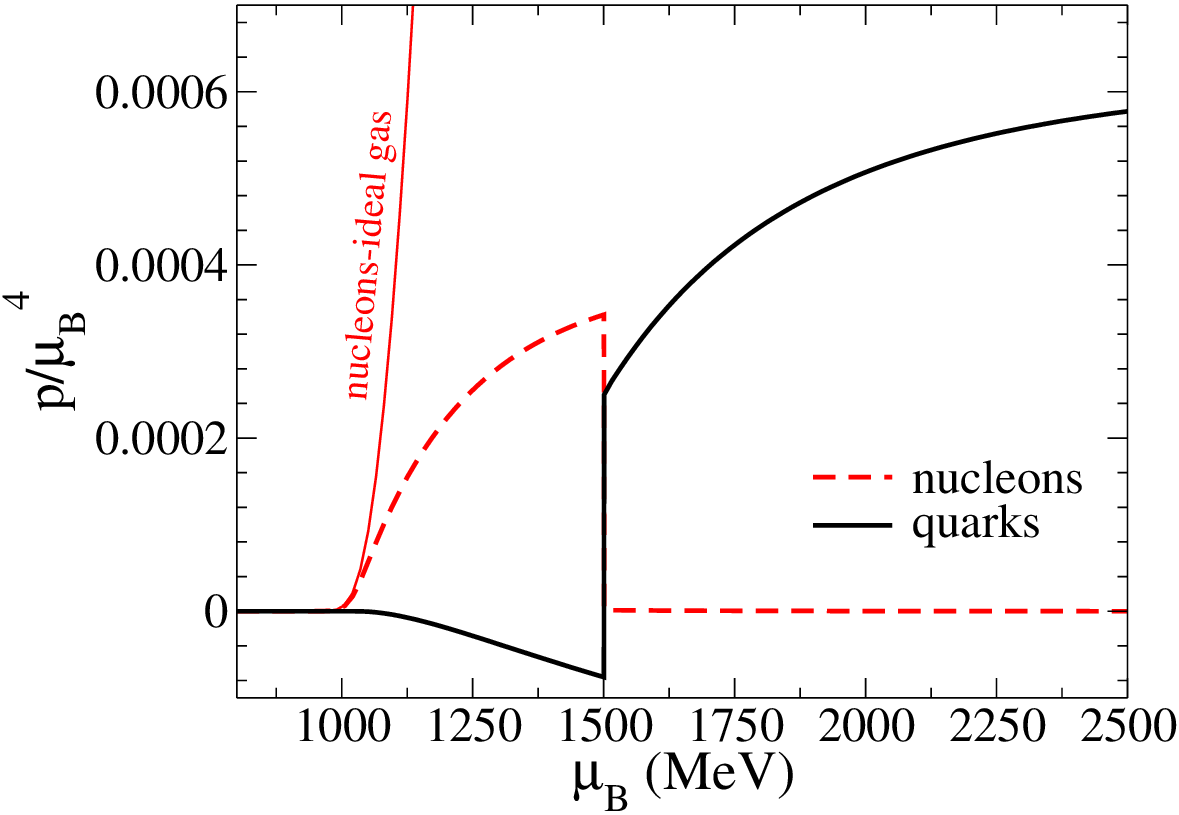}
\caption{Normalized partial pressures of the nucleon and the quark sectors as a function of $\mu_B$ in the simplified quark-meson-nucleon model.
The ideal nucleon pressure is plotted for comparison.}
\label{fig:pmjub}
\end{center}
\end{figure}
On Fig.~\ref{fig:pmjub} we plot the partial pressures as a function of $\mu_B$.
The nucleon pressure, given by the thick dashed red line, shows significant deviations from its ideal gas formula, shown by a thin red line.
In particular, at $\mu_B=1515$ MeV the nucleon pressure suddenly drops to zero value.
The pressure arising from the quarks at first turns to negative values.
The reason behind this is the increase of the $b$ field in the region $\mu_B =1000-1515$ MeV.
However, we stress that the total pressure of the system is always non-negative and continuous.
After $\mu_B = 1515$ MeV, nucleons disappear and the quark pressure becomes positive.
Moreover, since quarks become the dominant degrees of freedom in the system, the value $\mu_B = 1515$ MeV marks the deconfinement transition point for this simplified model.

\end{document}